# Localizing Faults in Cloud Systems

Leonardo Mariani*, Cristina Monni†, Mauro Pezzé*†, Oliviero Riganelli* and Rui Xin†
*Università degli studi di Milano Bicocca
Viale Sarca 336, Milano, Italy 20126
Email: {leonardo.mariani, mauro.pezze, oliviero.riganelli}@unimib.it
†Università della Svizzera Italiana (USI)
Via Buffi 13, Lugano, Switzerland 6900
Email: {cristina.monni, mauro.pezze, rui.xin}@usi.ch

*Abstract*—By leveraging large clusters of commodity hardware, the Cloud offers great opportunities to optimize the operative costs of software systems, but impacts significantly on the reliability of software applications. The lack of control of applications over Cloud execution environments largely limits the applicability of state-of-the-art approaches that address reliability issues by relying on heavyweight training with injected faults.

In this paper, we propose *LOUD*, a lightweight fault localization approach that relies on positive training only, and can thus operate within the constraints of Cloud systems. *LOUD* relies on machine learning and graph theory. It trains machine learning models with correct executions only, and compensates the inaccuracy that derives from training with positive samples, by elaborating the outcome of machine learning techniques with graph theory algorithms. The experimental results reported in this paper confirm that *LOUD* can localize faults with high precision, by relying only on a lightweight positive training.

## I. INTRODUCTION

Runtime failures are unavoidable in complex systems, and challenge the reliability of software applications [11]. Runtime failures are particularly hard to prevent in the cloud [48], since cloud systems rely on commodity hardware that reduces the overall system reliability, change rapidly to match evolving and sometime conflicting requirements, and suffer from confidentiality and visibility issues that may derive from the diffuse ownership of the system components [2]. The growing role of cloud systems in business-critical applications, like virtual retail stores [10], and the trend towards moving reliability-critical applications to the cloud, like the next generation telecom infrastructures [18], introduce strict reliability requirements that further exacerbate the problem.

The substantive characteristics of the cloud and the stringent reliability requirements of software applications running in the cloud, hereafter *cloud applications*, can be addressed at runtime with approaches that predict and localize faults [15, 16, 8, 9, 20, 19, 35, 29, 38, 41, 43, 44, 45, 48, 49] to trigger either automatic or manual recovery actions.

State-of-the-art fault localization approaches rely on long training sessions based on injected faults, which is an expensive and seldom-applicable practice, since it is hard to inject multiple classes of faults in several machines with different ownerships, and repeat this process while the system evolves.

In this paper we propose *LOUD*, an approach to accurately localize faulty components in cloud virtualized environments. *LOUD* locates faults as precisely as competing approaches [51, 9, 38], and is much more applicable in practice than competing approaches, since it exploits models trained under normal execution conditions only.

*LOUD* combines *machine learning* with *graph centrality* algorithms: It relies on Key Performance Indicators (KPIs), that is, metrics commonly collected on the running systems; It exploits machine learning to both *detect anomalies* in KPIs and reveal *causal relationships* among them; It complements machine learning with algorithms based on *centrality indices* to *localize* the faulty resources responsible for generating and propagating anomalies. *LOUD* originally exploits the causal relationships among KPIs and centrality indices to identify the causes of the failures, thus compensating the imprecision of anomaly detection based on models trained with positive samples only that are notoriously prone to false positives.

*LOUD* can be paired with any state-of-the-art failure predictor to localize the likely faulty resources before observing the failure, to enable automatic and manual healing.

The results of the experiments reported in Section VI indicate that *LOUD* can locate faults with a precision comparable to state-of-the-art approaches that rely on models trained with injected faults, and with an accuracy that varies over time.

The main contributions of this paper are: (i) an approach for modelling temporal correlations among anomalous values of KPIs, (ii) a study of the relation among anomalous KPIs during faulty executions, (iii) a technique to detect anomalies that are most likely related to the fault, based on a set of graph centrality indices, (iv) the experimental evaluation of the proposed approach for locating different types of faults in the context of various kinds of faulty resources.

The paper is organized as follows. Section II overviews the *LOUD* approach. Sections III and IV present the core data analytics and localizer components, respectively. Section V discusses the methodology we used to evaluate the effectiveness of the proposed approach. Section VI illustrates the experimental results about the effectiveness of *LOUD*. Section VII discusses related work. Section VIII summarizes the main contribution of the paper.

## II. LOUD

*LOUD* is an online metric-driven fault localization technique, which analyses the dependencies among anomalous

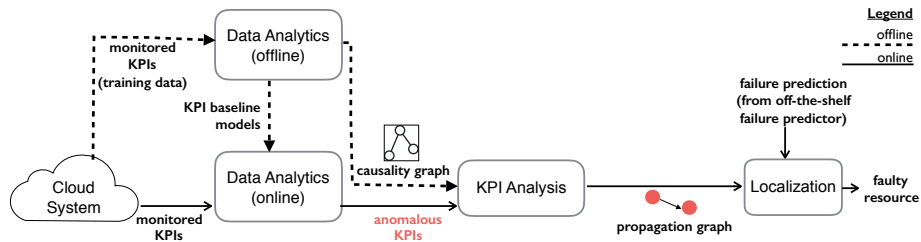

Fig. 1. The LOUD approach.

Key Performance Indicators (KPIs) commonly available in software systems at different abstraction levels to pinpoint the faulty resources that are likely responsible of future failures.

Figure 1 illustrates the offline and (subsequent) online phases that characterize the *LOUD* logical architecture. In the offline phase, *LOUD* trains a model that captures the normal system behavior. In the online phase, *LOUD* uses the model to pinpoint the faulty resources that are likely to be responsible of a possible failure.

The *LOUD Data Analytics offline phase* monitors the cloud system that operates in the field to model the system execution under normal conditions in the form of relations among *Key Performance Indicators (monitored KPIs)*, which are metrics collected on a regular basis from target resources of the system at different abstraction levels [12]. Examples of KPIs are the amount of occupied memory in a physical server, the CPU consumption of a virtual machine and the number of connections accepted by an application. These KPIs are sampled with probes that are deployed in the monitored resource [17, 3].

The *LOUD Data Analytics offline phase* produces a set of baseline models of the KPIs and a causality graph. The *KPI baseline models* represent the values of the KPIs that characterize normal behaviors. The *causality graph* models the causal dependencies among KPIs. Causal dependencies indicate the strength of the correlation between pairs of KPIs, that is, the extent to which changes in one KPI can cause changes in another KPI. The nodes in the causality graph correspond to KPIs, and the weighted edges indicate the causal relationships among KPIs. *LOUD* computes the baseline models and the causality graph using data collected when monitoring normal executions. *LOUD* exploits both the KPI baseline models and the causality graph in the online phase to localize faulty resources.

During the online operations, *LOUD* regularly monitors and verifies the KPIs against the baseline models (*Data Analytics–online*), creates the propagation graph, and localizes the faulty resource (*Localization*). *LOUD* executes the online as well as offline activities on an independent machine to avoid side effects on the production system, which is monitored with lightweight monitoring probes.

The *LOUD Data Analytics online phase* relies on *IBM ITOA-PI* [17], to identify the anomalous KPIs, that is, KPI values that violate the KPI baseline models. The *LOUD Localization* phase combines the set of anomalous KPIs with the causality graph to derive the *propagation graph*, which is the subgraph of the causality graph with anomalous KPIs only. Intuitively, it models the mutual influence among the anomalous KPIs. As the set of anomalies changes over time, the propagation graph also changes at every timestamp.

*LOUD* is designed to operate in cascade to a failure predictor. By relying solely on the prediction of a possible failure, *LOUD* can be integrated with any off-the-shelf predictor [32, 13, 38, 34, 47, 50, 46]. When triggered by a failure prediction, the *LOUD Localization phase* analyzes the propagation graphs produced at runtime, and identifies the faulty resource that corresponds to the likely root cause of the failure.

*LOUD* localizes faults under the hypothesis that the anomalous KPIs related to a fault are highly correlated and form a connected subgraph of the propagation graph. In particular, *LOUD* assumes that the incorrect behavior of a faulty resource is likely to result in incorrect behavior of neighbor resources, which are resources that interact with the faulty one either directly or indirectly. Intuitively, a misbehaving resource, such as a client that generates an unusually high load of requests or a virtual machine that consumes an unusually high amount of memory, usually affects related resources, such as the server application that processes the requests issued by the client or the host machine that runs the VM. Based on this assumption, *LOUD* exploits graph centrality indices [24, 27, 39] to identify the anomalous KPIs that best characterize the root cause of a predicted failure.

Since KPIs are metrics collected on specific resources, the KPIs with the highest centrality scores likely indicate the faulty resources that might be responsible for the predicted failure. The empirical results reported in Section VI indicate that the *LOUD* strategy based on centrality indices can effectively localize several classes of relevant faults in resources of different types.

In the next sections we describe the Data Analytics, KPI Analysis and Localization phases.

### III. DATA ANALYTICS AND KPI ANALYSIS

The *Data Analytics* component is composed of an offline and an online engine. The *offline* engine builds both a baseline model for each KPI, which captures the legal behavior of the KPI, and a causality graph, which captures causal relations between KPIs. The *online* engine identifies anomalies, that is, KPIs that violate their baseline models. Both the offline and online data analytics engines can be implemented with off-the-shelf IT operations analytics tool suites. We implement the *LOUD* Data Analytics engine with *IBM ITOA-PI*, a reliable product that provides advanced analytics technologies. To make the paper self-contained, we provide essential information about *IBM ITOA-PI*. The interested can refer to the documentation [17] for additional details.

The Data Analytics component elaborates data about KPIs provided in the form of time ordered sequences of data points. A KPI is a pair $\langle M, R \rangle$, where $M$ is a metric and $R$ is the monitored resource. The metrics capture measurable aspects of the behavior of the monitored system, for example memory consumption, number of requests served per minute, and so on. The resource can be any element of a system, for instance a host, a virtual machine, and a specific application.

Table I shows the sample case of a same metric, *Busy_CPU*, which indicates the percentage of CPU usage, collected from two sample resources, the virtual machines *bono* and *ellis*.

TABLE I
EXAMPLE DATA COLLECTED FOR TWO KPIS

| **Timestamps** | $\langle Busy\_CPU, Bono \rangle$ | $\langle Busy\_CPU, Ellis \rangle$ |
|---|---|---|
| 1440765146 | 0.348 | 0.137 |
| 1440765506 | 0.447 | 0.164 |
| 1440765866 | 0.539 | 0.137 |

*IBM ITOA-PI* produces a baseline model for each KPI and a causality graph for the whole system from the KPI data collected during an initial training phase that spans across multiple weeks of runtime and a constant retraining phase which takes place when new KPI data are available. The baseline model represents information about the average values and the acceptable variations over time of a KPI, by considering data seasonality. A KPI value monitored after the training phase is reported as anomalous if outside the acceptable variations coded in the baseline model or if the causal relationships represented in the graph have broken.

*IBM ITOA-PI* updates the model periodically at runtime to prevent false negatives and reduce any bias that might derive from incorrect behaviour occurring during the learning phase.

The causality graph is a directed weighted graph where nodes correspond to KPIs and edges represent causal relations among KPIs. A directed edge with a weight $w$ between nodes $n_a$ and $n_b$ that correspond to the KPIs $a$ and $b$, respectively, indicates that $a$ is correlated with $b$ according to the Granger statistical test with a probability $w$. The Granger causality test is specifically designed to capture correlation among time series data [14, 1].

The *KPI Analysis* phase exploits the causality graph and the set of anomalous KPIs to produce the *propagation graph*, which is a model that represents the anomalous KPIs and their interaction. Since the set of anomalous KPIs changes at each timestamp, the propagation graph is also different at each timestamp.

The propagation graph can be derived from the causality graph by preserving only the nodes corresponding to anomalous KPIs and their direct connections. More formally, given a set of anomalous KPI $K_A$ and a causality graph $(K, E)$, where $K$ is the set of KPIs and $e = (n_1, w, n_2) \in E$ is the set of weighted edges between KPIs, the propagation graph is the maximal connected subgraph $(K_A, E')$, with $E' = \{(n_1, w, n_2) \in E | n_1, n_2 \in K_A\}$. Figure 2 shows a sample propagation graph (on the right hand side) that has been generated from the causality graph (on the left hand side) in our experiments.

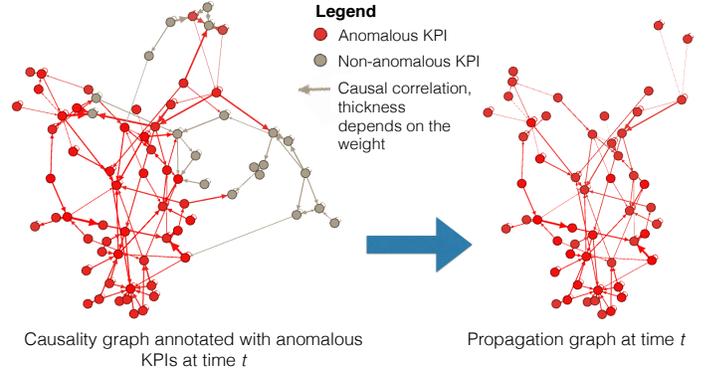

Causality graph annotated with anomalous KPIs at time *t* → Propagation graph at time *t*

Fig. 2. A sample causality graph and a corresponding propagation graph

## IV. LOCALIZATION

When triggered by a failure prediction, the *LOUD Localization phase* identifies the likely faulty resource by first ranking the KPIs whose values are detected as anomalous, and then selecting the faulty resource based on the ranking. The *Localization phase* is grounded on the observation that, once activated, faults cause increasingly many anomalies that spread both within the faulty resource and to resources either directly or indirectly interconnected to the faulty one. For instance, when a memory leak in a process exhausts the memory of the server, it may prevent the process generating responses to incoming requests, thus propagating to processes that share the memory with the faulty process, such as the virtual and host machines that run the process, as well as to other client processes that do not receive the expected responses, thus spreading anomalies within the same and to correlated resources.

The problem of identifying the fault location can be formulated as the problem of locating the nodes that originated the spreading of the anomalies represented in the series of propagation graphs built at runtime.

The problem cannot be solved by simply selecting the node that becomes anomalous first in the time series of propagation graphs because (i) the *data analytics component* identifies anomalies both in normal and in faulty executions, so it is hard to distinguish between a "root" anomaly and a false alarm, (ii) often the first anomaly does not correspond to the actual fault, but may correspond to an indirect effect not always easily traceable to the root fault.

*LOUD* works based on the assumption that the faulty resource is likely to generate an increasing number of strongly correlated anomalies over time during a faulty execution. This assumption comes from experimental evaluation over all the performed experiments both on faulty and failure-free executions.

The propagation graph is a powerful model to study the spread and the impact of anomalies across resources since its edges capture dependencies among KPIs. For instance, the propagation graph may capture the indirect correlation between the KPI that corresponds to the number of packets sent by a resource and the KPI that corresponds to the memory consumption of a machine that does not even receive these packets, but is indirectly impacted by the generated traffic.

*LOUD* identifies the nodes in the propagation graph that correspond the faulty resource, assigning scores through centrality indices, which are commonly used to identify the relevant nodes in weighted graphs with respect to the connectivity among nodes. The nodes in the propagation graph with the highest centrality scores identify the anomalous KPIs most likely related to the faulty resource. Thus we have a series of highly scored KPIs, one for each timestamp.

We considered different centrality indices, by focusing on the ones that take into account the global influence of a single node on the whole graph. For this reason, we discarded:

(i) the *degree centrality* and its generalizations because they do not consider how anomalies can spread on a graph,

(ii) the *betweenness centrality* and *closeness centrality* indices since they are based on shortest paths and thus do not take into account the global structure of the graph, which is important for the diffusion of anomalies.

We considered *eigenvector centrality* and its generalizations, *non-backtracking centrality*, *HITS algorithm for hubs and authorities*, and *PageRank*. These indices meet our requirements because they assign the highest score to the node with the highest information flow, and take into account the directions and weights of the links, and the presence of noise in the spread of the information within the graph.

Below, we introduce the centrality indices that we exploited for fault localization. We define all the indices referring to a same representation of the propagation graph based on the adjacency matrix $W$, where $W_{ij}$ is the value of the weight $w(i, j)$ of the edge from node $i$ to node $j$. The value is 0 if there is no edge between node $i$ and node $j$.

The *eigenvector centrality* is a vector **c** whose $i^{th}$ component $c_i$ represents the score of the node $i$ [39]. In a graph with $n$ nodes, the score $c_i$ is

$$c_i = \mu \sum_{j=1}^{n} W_{ij} c_j \quad (1)$$

that is proportional to the sum of the weights of the edges that connect the node $i$ to its neighbors, where $\mu$ is a proportionality constant that derives from the matrix of the eigenvalues. The eigenvector centrality score can be computed iteratively from Equation 1, and the solution is unique for a strongly connected graph with non-negative weights, as the case of propagation graphs.

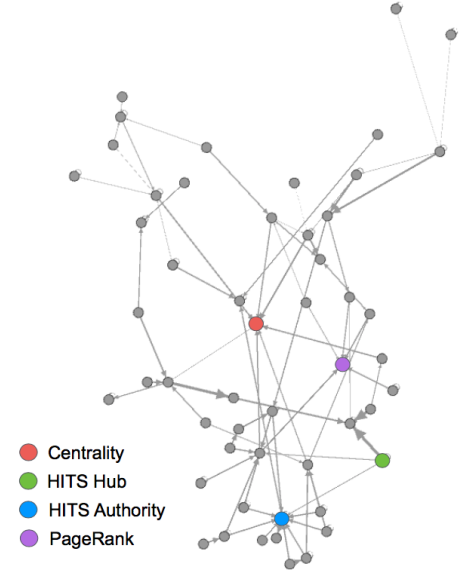

Fig. 3. Sample rankings for the nodes of the propagation graph of Figure 2

The eigenvector centrality may produce misleading results when the graph contains very few nodes that act as strong hubs and create a "condensation" or "winner-takes-it-all" phenomenon [27]. This is because the eigenvector centrality assigns to such nodes a much higher score than to most of the nodes, thus hiding the relevance of the other nodes which get a score close to zero, even if they might be related to the faulty resource. In our case, this phenomenon may potentially bias the localization and cause false positives.

The *non-backtracking centrality* is a variation of the eigenvector centrality that produces more accurate centrality scores in the presence of a condensation effect. Intuitively, the non-backtracking centrality scores a node $i$ as the sum of the scores of its neighbors computed in the absence of the node $i$ itself [27].

The non-backtracking centrality can be computed from equation (1) by substituting the adjacency matrix with the non-backtracking matrix $B$, defined as $B_{ij} = W_{ij}$ iff $(i, j)$ forms a non-backtracking path of length 2. A non-backtracking path is a directed path in which no edge is the inverse of the preceding edge [31].

The assumption behind the eigenvector and non-backtracking centrality is that a node is important if it is attached to many edges coming from many other important nodes. In the propagation graph, these nodes represent KPIs which are more influenced by the spread of anomalies.

*The Hyperlink Induced Topic Search* (HITS) algorithm assigns a high score to nodes that are linked with other highly scored nodes [23]. This intuitively corresponds to identifying the KPIs that have a strong influence on the behavior of the system, and thus the corresponding resources. These relevant nodes might be of two kinds: *authorities*, which are the nodes that are the destination of edges from highly ranked nodes, and *hubs*, which are the nodes with many outgoing edges leading to highly ranked nodes.

The equations to score nodes as authorities ($a_i$) or hubs ($h_i$) are:

$$\sum_{j=1}^{n} W_{ij}\, a_j = \delta\, h_i, \qquad \sum_{k=1}^{n} W_{ik}\, h_k = \nu\, a_j$$

where $\delta$ and $\nu$ are proportionality constants related to eigenvalues of $W$. The HITS algorithm combines the two equations to compute iteratively a hub score and an authority score for each node in the graph.

*PageRank* [24] scores nodes according to both the number of incoming edges and the probability of anomalies to randomly spread through the graph (*teleportation*). The PageRank score $r_i$ of node $i$ is computed iteratively from the score $r_j$ of the nodes $j$ that are sources of edges directed to $i$:

$$r_i = \sum_{j \in \mathcal{N}_{in}(i)} W_{ij}\, r_j + \frac{1-\alpha}{n} \qquad (2)$$

where $\mathcal{N}_{in}(i)$ is the set of nodes $j$ that are sources of edges directed to $i$, and $\alpha$ is the teleportation probability (*LOUD* uses $\alpha = 0.85$). The interested reader can refer to the excellent description of Langville et al. [24] for a detailed survey on the HITS algorithm and PageRank.

Figure 3 exemplifies the effects of the different centrality indices on the propagation graph of Figure 2. The Figure shows in different colors the nodes with the highest score for each algorithm. The red node is the top ranked node of both eigenvector and non-backtracking centrality, which means that the centrality index is evenly distributed through many important nodes in the graph, and the variations of the non-backtracking centrality have a low impact on the scores. The Figure illustrates the dual role of hubs and authorities: there is a directed edge from the top HITS Hub node to the top HITS Authority node. The role of the purple highest PageRank node is similar to the role of the red highest centrality node, since both are destination of highly weighted edges. The two nodes are different due to the impact of the PageRank teleportation.

*LOUD* computes the four centrality indices with the power iteration algorithm that computes the indices with an increasingly better approximation at each iteration, and can thus be interrupted and later resumed at any time to improve the precision of the localization.

*LOUD* identifies the likely faulty resource as the most frequent resource in the top 20 KPIs according to the selected centrality index, and does not suggest a fault location in the absence of a most frequent resource in the set. The intuition is that a faulty resource is likely to show an anomalous behavior for multiple KPIs, which in turn are likely to affect several other resources, and thus their KPIs, of the system. As a consequence a faulty resource is likely to be present with multiple KPIs in the top part of the ranking.

## V. EVALUATION

In this section, we describe the experimental setting for evaluating *LOUD*: We introduce the research questions (Section V-A), the testbed that implements the cloud-based system that we use in the evaluation (Section V-B), the process for executing the experiments and collecting the data from the testbed (Section V-C), and the set of faults that we seeded in the system for evaluating *LOUD* (Section V-D).

### A. Research Questions

In the experiments, we addressed three research questions that qualify the effectiveness and overhead of *LOUD*:

**RQ1:** *Does the choice of centrality index impact on the precision of* LOUD *fault localization?*
As discussed in Section IV, *LOUD* implements five centrality-based indices that are potentially suitable for fault localization. We experimentally compared the effectiveness of the different indices, to evaluate their impact on fault localization.

**RQ2:** *Do the type of the faulty resource, the type of the fault and the activation pattern impact on the effectiveness of* LOUD *localization?*
We evaluated the accuracy of *LOUD* with different kinds of faults injected in various types of resources with distinct activation patterns to evaluate the impact of these factors on *LOUD* precision. We executed the experiments with the PageRank centrality index, which the experiments for RQ1 indicate as the most effective index for the *LOUD* localization.

**RQ3** *What is the overhead of* LOUD *on the cloud environment?*
*LOUD* is a lightweight approach that isolates the resource consuming activities on a dedicated server, and deploys only probes in the cloud environment. Thus, probes are the only *LOUD* elements that may impact on the cloud system. We evaluated the impact of the *LOUD* probes by comparing the execution time of the monitored resources with and without active probes.

### B. Testbed

We executed the experiments on a cloud environment running ClearWater,[1] an open source implementation of an IP Multimedia Subsystem. The cluster running ClearWater consists of 8 server machines that share a local subnet. Each machine runs Ubuntu 14.04 LTS. We installed OpenStack Icehouse[2] to manage the cluster. We configured the cluster with *six compute nodes* that run the KVM hypervisor[3] and host the Virtual Machine (VM) instances that run ClearWater,

---

[1]Project Clearwater. IMS in the Cloud. http://www.projectclearwater.org. Last access: oct 2017
[2]The OpenStack Project. Open source software for creating private and public clouds. https://www.openstack.org/. Last access: oct 2017
[3]The KVM Project. Kernel Based Virtual Machine. https://www.linux-kvm.org/. Last access: oct 2017.

*a controller node* that offers VM managing services, from the basic ones, like creation, deletion and reboot, to more advanced ones, like identity services, image service and dashboard, and *a network node* that routes virtual networks through NAT.

Table II shows the hardware and software configuration of each machine.

TABLE II
INFRASTRUCTURE AND PLATFORM CONFIGURATION

| Role | Controller | Network | Compute (x2) | Compute (x4) |
|---|---|---|---|---|
| **CPU** | *Intel(R) Core(TM)2 Quad CPU Q9650* <br> *(12M Cache, 3.00 GHz, 1333 MHz FSB)* | | | |
| **RAM** | 4 GB | 4 GB | 8 GB | 4 GB |
| **Disk** | 250 GB SATA hard disk | | | |
| **NIC** | Intel(R) 82545EM Gigabit Ethernet Controller | | | |

Clearwater [6] exploits six virtual machines to provide messaging, voice and video communication using the SIP protocol. In particular: *Bono* serves as entry point for users connections, *Sprout* routes user requests forwarded by Bono and checks authentication, *Homestead* provides the user authentication and profile data, which are stored in a database for Sprout to inquire through a web service interface, *Homer* stores MMTEL service settings of users in the form of standard XML, *Ralf* provides offline billing features by interacting with Bono and Sprout, *Ellis* runs as a web sever for users to register and manage personal profile and settings.

Figure 4 illustrates the logical architecture of the system, indicating dependencies across layers.

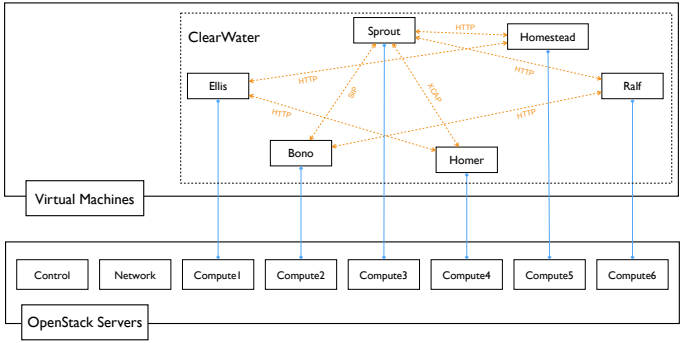

Fig. 4. Reference Logical Architecture

### C. Data Generation and Collection

We executed our testbed with a workload implemented with the SIPp open source SIP traffic generator[4]. We shaped the number of users and calls in the workload based on *week* and *day* patterns: *within each week*, the number of users in working days is higher than in weekends, *with each a day*, the number of users grows at daytime and decreases at nighttime, with peaks at 9am and 7pm.

We based the *LOUD* monitoring on KPIs collected from the both the operating system and ClearWater. At the *operating system* level, *LOUD* uses a set of monitoring probes that collect 25 KPIs on the status of both the machines and the communication interfaces. *LOUD* implements the monitoring probes with common Linux monitoring tools such as the iostat, sar, vmstat, free, ps, ping, and the Psutil Python library. *LOUD* monitors the operating systems running on both the host and the virtual machines, and the machines that host the compute nodes. *LOUD* collects a total of 300 KPIs at the operating system level for the 6 compute nodes and 6 virtual machines. At the *Application* level, *LOUD* uses the SNMPv2c [5] monitoring service for ClearWater, which provides a total of 162 KPIs for the 6 ClearWater machines.

The *LOUD* monitor samples the 462 KPIs every minute, and sends the collected data to the analysis server that runs the *LOUD* approach. In our setup, the analysis server runs Red Hat Enterprise Linux Server 6.3 with an Intel(R) Core (TM) 2 Quad Q9650 processor at 3GHz frequency and 16GB RAM.

We trained *LOUD* offline with data collected by executing the testbed for 2 weeks under normal conditions, as recommended by *IBM ITOA-PI*. We used the default setup for *IBM ITOA-PI*, which aggregates and analyzes data in time intervals of 5 minutes.

### D. Investigated Faults

We simulated faulty scenarios using fault injection techniques. We injected fault types commonly used in the evaluation of state-of-the-art fault localization approaches [38, 36, 16, 22]: packet loss, memory leak and CPU hog. For each fault we considered different severity growth patterns: (i) *linear pattern,* the fault is triggered with a same frequency over time, (ii) *exponential pattern,* the fault is activated with a frequency that increases exponentially, resulting in a shorter time to failure, (iii) *random pattern,* the fault is activated randomly over time. We injected the faults by adapting the ChaosMonkey open-source failure generator[5] to enable the various types of fault severity growths.

We injected each type of fault in Bono, Sprout and Homestead, which are the three core components of ClearWater. To increase the generality of the results, we repeated the injection process four times for each type of fault. This produces a total of 108 experimented cases: 3 types of faults (packet loss, memory leak, CPU hog) injected in 3 different resources (Bono, Sprout, Homestead) with 3 different activation patterns repeated four times.

### E. Quality Metrics

We measure the effectiveness of *LOUD* (RQ1 and RQ2) by computing precision, recall, and F1-score of the localization at each timestamp after the activation of the fault. In particular, we define true/false positive/negatives as follows:

- *true positive (TP) at time $t$:* the *LOUD* localization produced at time $t$ is correct,
- *false positive (FP) at time $t$:* the *LOUD* localization produced at time $t$ is wrong,
- *false negative (FN) at time $t$:* *LOUD* cannot identify a most frequent resource in the top 20 anomalies, and thus does not produce a localization at time $t$.

---

[4]Gayraud Richard and Jacques Olivier. SIPp. http://sipp.sourceforge.net. Last access: may 2015.

[5]Cory Bennett. Chaos Monkey. https://github.com/Netflix/SimianArmy/wiki/Chaos-Monkey. Last access: oct 2017.

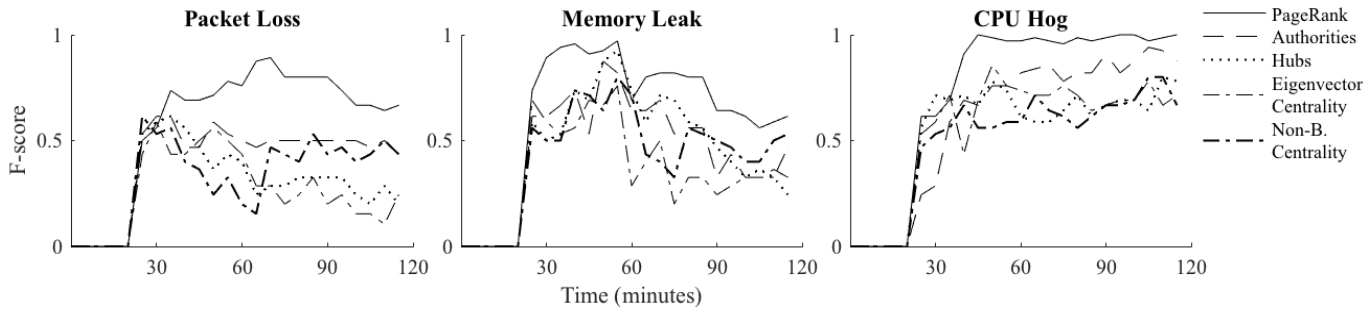

Fig. 5. F1-score of each technique per fault type

We compute the precision as the rate of correct localizations out of the total set of localizations produced at time $t$: $\frac{TP}{TP+FP}$. We compute the recall as the rate of correct localizations out of the total set of localizations that should have been generated at time $t$: $\frac{TP}{TP+FN}$. We compute the F1-score as the harmonic average between precision and recall: $2\frac{precision*recall}{precision+recall}$.

We measure the intrusiveness of the probes by computing the overhead of the probes (RQ3).

## VI. EXPERIMENTAL RESULTS

In this section, we discuss the results of our experiments organized according to the research questions presented above.

### A. RQ1: Does the choice of centrality index impact on the precision of LOUD fault localization?

We evaluate the impact of the centrality index by measuring the effectiveness of *LOUD* in localising faults, when executing *LOUD* with the different indices for various types of injected faults. Figure 5 shows how the effectiveness of the localization changes over time for the five centrality indices that we identified in Section IV, eigenvector centrality, non-backtracking centrality, HITS algorithm for hubs and authorities, and PageRank, and for the three classes of faults that we considered, packet loss, memory leak, and CPU hog.

The diagrams reported in this and all the figures of the paper show the F1-score computed for experiments starting from a fault injected at time 0 and lasting 120 minutes. The null value of F1-score for the first 20 minutes of the experiments derives from *IBM ITOA-PI*, which does not compute anomalies for the first 20 minutes of execution. In all the experiments, the cloud system failed after the 120 minutes time interval shown in the plots.

Figure 5 allows us to draw some considerations:

*LOUD performs better with PageRank than with any other considered index:* *LOUD* with PageRank dominates every other algorithm for all fault types through most of the observed time interval. This indicates that PageRank, and the concept of teleportation that encodes the probability of anomalies to spread non-uniformly across the graph, as it may happen in real scenarios, well captures how anomalies can spread in cloud systems.

*The effectiveness of LOUD localization depends on the fault type:* The effectiveness of fault localization strongly depends on the kind of injected fault. Indeed, anomalies spread according to significantly different patterns for different fault types. Although the centrality indices present different effectiveness for different fault types, the relative difficulty to localize a fault does not depend on the specific centrality index. Packet loss faults are the hardest faults to localize for all the indices, likely because the effect of network problems easily and quickly spreads through the system resources, in a way that is difficult to trace back to the source node. Memory leaks are also challenging but definitely easier to localize than packet loss faults. CPU hogs spread with patterns that are the easiest to trace back to the root cause among the considered faults for all algorithms.

*The effectiveness of the LOUD localization does not always improve with time:* The results do not confirm the intuition that the quality of the localization improves over time, due to the increased evidence of both the failures and their causes over time. CPU hogs are the only type of faults with an increasing quality of the localization over time. This is likely due to the increasingly stronger impact of the CPU utilization on the affected resource compared to the other resources of the system. In the presence of both packet loss and memory leaks, the quality of the localization first increases and then decreases, with some perturbation in the intermediate phase. This suggests that in an initial phase the anomalies incrementally produced by a fault accumulate in a way that facilitates the localization task, but at some point the spreading is so extensive that it gets confused within the noise of the system, that is, with the anomalies that are regularly produced by the system despite the presence of faults.

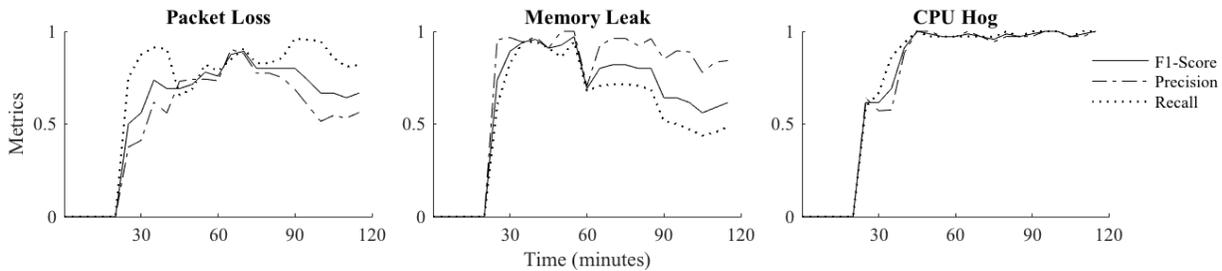

Fig. 6. F1-score, precision and recall of PageRank per fault type

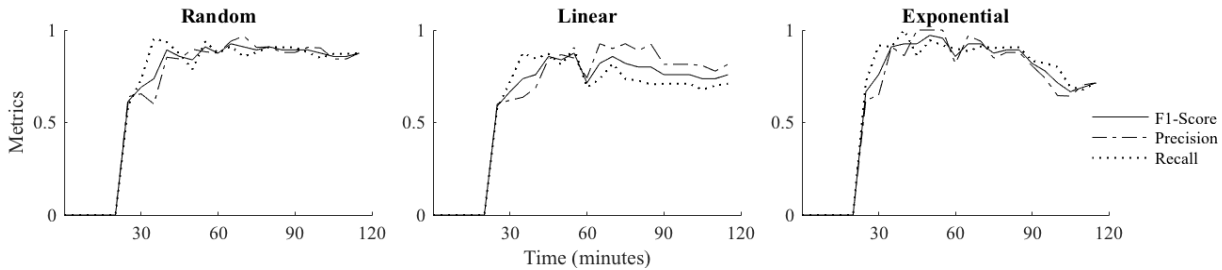

Fig. 7. F1-score, precision and recall of PageRank per activation pattern

Having identified PageRank as the index that leads to the best performance compared to the other centrality indices, we conducted all the other experiments with PageRank.

*B. RQ2: Do the type of the faulty resource, the type of the fault and the activation pattern impact on the effectiveness of* LOUD *localization?*

We investigate the effectiveness of the *LOUD* localization with respect to three key dimensions: fault types, fault activation patterns, and faulty resources.

Figure 6 shows F1-score, precision, and recall of *LOUD* with packet loss, memory leak and CPU hog injected faults. The precision and recall follow some interesting and complementary trends for the different classes of faults.

The localization of packet loss faults reaches a high F1-score only after a long time slack (about 65 minutes after the fault injection, and 40 minutes after the first *IBM ITOA-PI* anomaly), and does not stabilize, thus the localization often fails in identifying the faulty resource, as witnessed by a recall higher than precision. The localization of packet loss faults is precise only in a fairly short time interval (about 10 minutes) in terms of F1-score.

*LOUD* addresses well memory leak faults despite an unstable F1-score, as witnessed by an always high precision with a drop of the recall when far from the fault activation. Thus, *LOUD* may fail in identifying the fault location, but it identifies it precisely when it succeeds.

*LOUD* addresses well CPU hogs both in terms of precision and recall.

Figure 7 shows F1-score, precision and recall for the different activation patterns averaged over fault types. *LOUD* shows a similar trend for all the activation patterns, thus suggesting a reasonable independence of the effectiveness of the technique with respect to the growth rate.

The effectiveness of *LOUD* remains stable over time for both the linear and random activation patterns, and slightly decreases in the long term for the exponential pattern. This is likely due to the rapid increase and spread of anomalies that impact on the effectiveness of the localization algorithm.

Figure 8 shows the F1-score, precision and recall per resource, and indicates that *LOUD* is precise for all the resources, slightly less stable for Homestead. A careful inspection of the results indicates that the localization phase sometime erroneously identifies Sprout as the faulty resource instead of Homestead. The architecture of the testbed discussed in Section V-B shows that Homestead and Sprout are highly interacting. Since *LOUD* locates well faults for other highly interacting resources, for instance Bono and Sprout, the relatively lower effectiveness for Homestead may depend on the specific characteristics of the interaction. In our prototype setting, the KPIs are unevenly distributed among Sprout and Homestead: 92 KPIs for Sprout, 71 KPIs for Homestead. Since a fault in Homestead is likely to impact on both resources, the smaller number of KPIs for Homestead than Sprout may bias the fault localization in favour of Sprout. The results might be likely improved by either collecting a set of evenly distributed KPIs or introducing a normalization strategy that takes into account the number of KPIs extracted from each resource. Both studies are part of our current research plan.

*C. RQ3: What is the overhead of* LOUD *on the cloud environment?*

The *LOUD* overhead on the production system derives only from the monitoring probes, which collect and send data to the analysis node running *LOUD*. *LOUD* can be executed either on a separated node of the cloud system or on a separate machine outside of the cloud environment. In the former case, the cloud infrastructure has to allocate enough resources to support the execution of the analysis node, in the latter case, *LOUD* does not require additional resources from the cloud infrastructure. In both cases, *LOUD* does not introduce significant overhead on the monitored resources: In our experiments, we measured an average increase of 2.63% on CPU usage and 1.91% on memory usage of the monitored resources, which is a negligible overhead for cloud services.

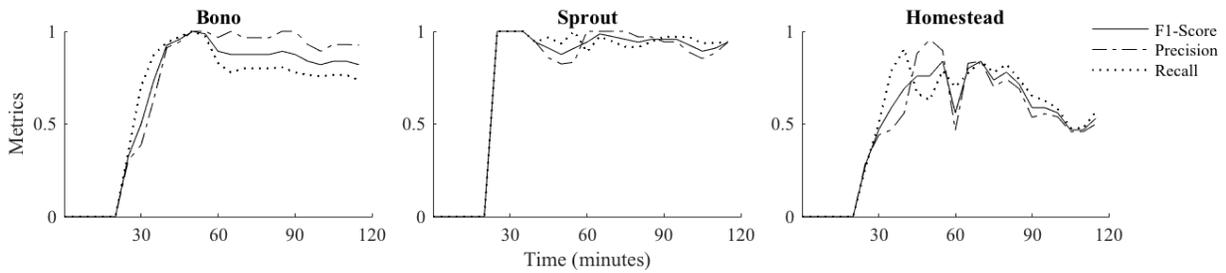

Fig. 8. F1-score, precision and recall of PageRank per resource

Overall, *LOUD* demonstrates to be effective in localizing faults in the cloud, with an effectiveness that depends on the kind of fault to be localized, but is largely independent from the fault activation pattern. In particular, *LOUD* is very effective with CPU hogs (the *LOUD* localization is precise, quick and constantly available after the activation of the fault), memory leaks (the *LOUD* localization is always precise, although sometime hindered by the noise in the anomalies), and useful with packet loss (the *LOUD* localization is precise, at least in some time interval after the fault activation). These results are extremely good, since the effectiveness of *LOUD* is in line with the effectiveness of competing approaches, but *LOUD* relies solely on training with normal executions, while competing approaches require long training sessions with injected faults [38]. Sessions with injected faults strongly limit the applicability of the approaches to large and evolving cloud systems that badly tolerate if at all artificially faulty sessions, and that cannot be replicated in the laboratory.

### D. Threats to Validity

The main threats to the validity of our results derive from the fault injection strategy, the configuration of the prototype and the type of faults that we considered in the experiments.

We mitigated the bias that may derive from the specific fault injection strategy that we adopted by both using ChaosMonkey, a freely available tool for fault injection, and considering multiple activation patterns to increase the generality.

The specific configuration of our prototype and of the set of faults that we investigated may limit the generality of our results. Building complex infrastructures and repeating the experiments with multiple testbeds is extremely expensive. We mitigate this bias by experimenting with ClearWater, an environment widely used in academia and industry since it is designed for scalable deployment in the cloud [38, 4].

The scalability of *LOUD* has not been tested, but it is potentially allowed by the nature of the algorithm, which is based on *IBM ITOA-PI*, a tool for big data analytics, and Pagerank which is by definition a scalable algorithm.

We experimented with a relatively small set of types of faults. Although *LOUD* does not depend on the specific class of faults, its effectiveness may depend on the faults as observed in the paper. For this reason, we report and discuss the results at the level of individual fault types.

## VII. RELATED WORK

So far, cloud-based fault localization [42, 25] focused on few fault types, the most common ones being performance faults, resource faults, and operational faults. *Performance faults* are faults that cause the unexpected degradation of the system performance, such as packet loss faults. Performance faults may be due to chronic problems [19, 21], task interference [30, 8], and latency increment [28, 15, 16, 37]. *Resource faults* [20, 7, 38, 33, 43] are faults that cause the incorrect allocation and use of system resources, such as memory leakages and CPU hogs. Resource faults may also derive from the excessive use and exhaustion of resources. *Operational faults* [9, 49, 26, 41, 29] are execution faults that impact on performance, and originate from deadlocks, infinite loops, unhandled exceptions and emergent behaviors. Our experimental evaluation with *LOUD* considered both performance and operational faults.

Fault localization techniques implement different analysis techniques: latency analysis, data analytics, machine learning, graph-based approaches and their combination.

*Latency Analysis* consists of identifying operations with anomalous latency to diagnose the possible faulty elements responsible of the anomalies. The localization can be at different granularity levels, from individual methods to network nodes. For example, CloudDiag [28] captures user requests, traces the execution time of the individual methods, and identifies the method calls that are likely responsible of an observed anomaly, based on the distribution of the latency time. Sometime the cause of a problem is not a single method but a sequence of method calls. This is why DARC [45] identifies the *root cause path*, which is a call-path that starts from a given function and includes the largest latency contributors to a given peak time. call-graph. Khanna et al. at define an architecture for the localization of faulty nodes and links by collecting network measurements [22]. *LOUD* is not limited to latency analysis but collects and analyzes a range of data, including the data necessary for performance analysis, as reported in the empirical evaluation.

*Data Analytics* exploits various kinds of statistical techniques to localize faults. PeerWatch [20] uses canonical correlation analysis (CCA) to identify the correlations between the characteristics of multiple application instances and localize faults. PerfCompass [8] collects data about the performance of system calls while virtual machines behave correctly, and analyzes the data collected in the presence of performance problems to classify the root cause of the fault as either external or internal. Johnsson et al. define an algorithm to determine the root cause of a performance degradation problem by using network level measurements and a graphical model of the network [19]. Herodotou et al. focus on data center failures [16], using active monitoring techniques, such as pings, to produce a ranked list of devices and links that are likely related to the root cause. While statistical techniques may capture the resources with a behavior correlated well with the observed failures, they cannot always trace a problem to its root cause, because they do not consider how misbehaviors propagate across the resources of the system, as centrality indices do in *LOUD*.

*Machine Learning* is another widely adopted solution to localize faults. UBL [7] leverages Self-Organizing Maps (SOM), a particular type of Artificial Neural Network, to capture emergent behaviors and unknown anomalies. Yuan et al. exploit machine learning to analyze system call information on Windows XP, and classify the information according to known faults [51]. POD-Diagnosis [49] deals with faults in sporadic operations, for instance, upgrade, and replacement, by extracting events from logs, detecting errors from events, and using a pre-constructed tree that encodes fault knowledge to identify the root causes. BCT [26] exploits behavioral models inferred from tests execution to identify the causes of software failures. Sauvanaud et al. [38] use Random Forest to detect Service Level Agreements (SLAs) violations and infer the root cause. While many techniques based on machine learning build their models involving prior characterization of the fault, through either fault injection or human knowledge, *LOUD* localizes faults by exploiting centrality indices and studying the propagation of the anomalies in the cloud. In this way, *LOUD* requires only data from normal executions and needs no predefined fault profile.

*Graph-based approaches* localize faults by using graph models derived from the network topology and dependencies between services. Sharma et al. propose an algorithm to localize the problematic metrics by using a dependency graph built with invariants among system metrics [40]. Gestalt [29] combines the features of existing algorithms to find the culprit component in a transaction failure, with the help of a dependency graph. Srikar et al. propose an adaptive algorithm to diagnose large-scale failures in computer networks [44], by analyzing a graph that represents the network topology. Graph-based approaches that refer to structural dependencies might miss the often relevant indirect impact that the behavior of a resource may have on another resource. *LOUD* analyzes a graph that represents causal relations between anomalous KPIs, thus referring to the monitored mutual impact among the resources in the system.

Some approaches combine different techniques similarly to *LOUD*, but focus on a single fault type. PerfScope [9] combines hierarchical clustering, outlier detection and finite state machine based matching, to address operational faults. Kahuna [43] combines clustering and latency analysis on Hadoop network resource faults. CloudPD [41] combines Hidden Markov Models (HMM), k-nearest-neighbor (kNN), and k-means clustering to build the performance model, and uses statistical correlations to identify anomalies and signature-based classification to identify operational faults. Pingmesh [15] combines data analytics and latency analysis to detect network faults.

## VIII. CONCLUSIONS

In this paper we present *LOUD*, an approach for localizing faults in cloud systems, which is based on graph centrality algorithms to efficiently locate faults with limited interference with the target system both during training and operational conditions. The *LOUD* approach improves over state-of-the-art techniques, by (i) relying on a training phase that requires only normal executions, thus supporting model training in the field without the relevant interferences that derive from the fault injection required by current techniques, (ii) running on an independent machine that works on metrics usually already collected in the field, and thus with negligible interference with the system, and (iii) locating faults with an efficiency in line with the best state-of-the-art approaches.

The unique *LOUD* feature of relying only on data from normal executions, jointly with the negligible execution overhead, and the good effectiveness make *LOUD* well suited for the many large over-running systems that cannot be suspended and sand-boxed to perform in-lab experiments.

## ACKNOWLEDGMENT

This work has been partially supported by the H2020 Learn project, which has been funded under the ERC Consolidator Grant 2014 program (ERC Grant Agreement n. 646867), the Italian Ministry of Education, University, and Research (MIUR) with the PRIN projects IDEAS (grant n. PRIN-2012E47TM2_006) and GAUSS (grant n. 2015KWREMX), by the H2020 NGPaaS project (grant n. 761557), and by IC Information Company AG.